\documentclass{article}

\usepackage{lastpage}
\usepackage{epsfig}
\usepackage{graphicx,amssymb,mathrsfs,amsmath}

\def\cdd{\mbox{\boldmath$\cdot$}~}
\makeatletter
\def\@oddfoot{\hfill}

\newcount\shumeicount
\def\setshumei#1#2#3{%
  \shumeicount=\count0
  \def\@oddhead{%
    \raise-5pt\hbox to0pt{\vrule width\hsize height 0pt depth 0.4pt\hss}\relax
    \ifnum \shumeicount=\count0
      \raise-7pt\hbox to0pt{\vrule width\hsize height 0pt depth 0.4pt\hss}\relax
      #1
    \else
      \ifodd\count0
        #2
      \else
        #3
       \fi
     \fi
  }%
}
\makeatother

\makeatletter
\def\@oddfoot{\hfill}
\newcount\shujiaocount
\def\setshujiao{%
  \shujiaocount=\count0
  \def\@oddfoot{%
      \ifodd\count0
      \else
      \fi
  }%
}
\makeatother

\def\biaoti#1#2#3#4{{
  \vspace*{0.3cm}
  \begin{flushleft} \Large\bf #1\end{flushleft}
  \vspace*{-0.2cm}
      \begin{flushleft}
      \bf #2
      \end{flushleft}
      \footnotetext{\hspace{-6mm} #3\\ #4}}}

\def\dshm#1#2#3#4
{\setshumei{ Jrl Syst Sci \& Complexity (#1) #2:
{\thepage--\pageref{LastPage}}\hfill}
            {\hfill {\small #3}\hfill\hbox to0pt{\hss\thepage}}
            {\hbox to0pt{\thepage\hss}\hfill {\small #4}\hfill
            }
            \setshujiao}
\def\drd#1
{{\vskip 1cm\small \begin{flushleft}
 #1 \\
\copyright 2006 Springer Science + Business Media, Inc.
\end{flushleft}}}
\def\dab#1#2{\noindent {{\small\bf Abstract~~}}{{\small #1}}
            \vskip 0.1cm
             \noindent{{\small\bf Key words~~}}{{\small #2}}
                      }

\def\dse#1{\vskip 0.6cm\noindent
        {\large\bf #1}
        \vskip 0.4cm}
\def\rf#1#2{\parindent=0pt\hangindent=0.6cm\hangafter=1\small
            \parbox[t]{0.6cm}{[#1]}#2\par}
\def\rfne{\vskip 0.5cm  \centerline{\bf References} \vskip 0.5cm
               \parindent 0pt}
  
\def\dsee#1{\vskip 0.3cm \noindent{\bf #1} \vskip 0.2cm}

\def\Rmn#1{\expandafter\uppercase\expandafter{\romannumeral #1}}
\usepackage{graphicx}

\newcommand{\vect}[1]{\mathrm{vec}(#1)}
 \newtheorem{theorem}{Theorem}
 \newtheorem{lemma}{Lemma}

 \newtheorem{algorithm}{Algorithm}

\begin{document}

\biaoti{OBTAINING EXACT INTERPOLATION MULTIVARIATE POLYNOMIAL BY
APPROXIMATION$^*$}{Yong FENG \cdd Jingzhong ZHANG \cdd Xiaolin QIN
\cdd Xun YUAN}{Yong FENG \\ {\it School of Computer Science and
Engineering, University of Electronic Science and Technology of
China,Chengdu} $610054$, {\it China}; {\it Laboratory for Automated
Reasoning and Programming, Chengdu Institute of Computer
Applications,Chinese Academy of Sciences,Chengdu} $610041$, {\it
China}.Email:yongfeng@casit.ac.cn.
\\Jingzhong ZHANG \\{\it Laboratory for Automated Reasoning and
Programming, Chengdu Institute of Computer Applications,Chinese
Academy of Sciences,Chengdu} $610041$, {\it China}.\\Xiaolin QIN\\
{\it Laboratory for Automated Reasoning and Programming, Chengdu
Institute of Computer Applications,Chinese Academy of
Sciences,Chengdu} $610041$, {\it China}.Email: qinxl811028@163.com.
\\Xun YUAN \\{\it Laboratory
for Automated Reasoning and Programming, Chengdu Institute of
Computer Applications,Chinese Academy of Sciences,Chengdu} $610041$,
{\it China}.} {$^*$This research is supported by China 973 Project
NKBRPC-2004CB318003, the Knowledge Innovation Program of the Chinese
Academy of Sciences KJCX2-YW-S02, and the National Natural Science
Foundation of China(Grant NO. 10771205).}

\drd{Received: 3 September 2008}{ }{}{}


\dshm{}{}{OBTAINING EXACT INTERPOLATION MULTIVARIATE POLYNOMIAL BY
APPROXIMATION}{YONG FENG $\cdot$ JINGZHONG ZHANG $\cdot$ XIAOLIN QIN
$\cdot$ XUN YUAN}

\dab{In some fields such as Mathematics Mechanization, automated
reasoning and Trustworthy Computing etc., exact results are needed.
Symbolic computations are used to obtain the exact results. Symbolic
computations are of high complexity. In order to improve the
situation, exactly interpolating methods are often proposed for the
exact results and approximate interpolating methods for the
approximate ones. In this paper, we study how to obtain exact
interpolation polynomial with rational coefficients by approximate
interpolating methods.} {Numerical approximate computation,
symbolic-numerical computation, continued fraction, multivariate
interpolation, Vandermonde determinant.}

\dse{1~~Introduction}

 Some fields such as automated reasoning and
trustworthy computing etc., need exact results, and symbolic
computations are used to obtain the exact results. Symbolic
computations are principally exact and stable. However, they have
the disadvantage of intermediate expression swell. Numerical
computations have not the problem, however only give approximate
results. In recent two decades, numerical methods are applied in the
field of symbolic computations. In 1985, Kaltofen presented an
algorithm for performing the absolute irreducible factorization, and
suggested to perform his algorithm by floating-point numbers, then
the factor obtained is an approximate one. After then, numerical
methods have been studied to get approximate factors of a
polynomial$^{[1-6]}$. In the meantime, numerical methods are applied
to get approximate greatest common divisors of approximate
polynomials$^{[7-10]}$, to compute functional
decompositions$^{[11]}$, to test primality$^{[12]}$ and to find
zeroes of a polynomial$^{[13]}$. In 2000, Corless et al. applied
numerical methods in implicitization of parametric curves, surfaces
and hypersurfaces$^{[14]}$, and the resulting implicit equation is
still an approximate one. In ${[15]}$,Ch$\grave{e}$ze et al.
discussed how to obtain an exact absolute polynomial factorization
from its approximate one, which only involves recovering an integral
coefficient from its approximation.

Interpolation methods as an efficient numerical method have been
proverbially used to compute resultants and determinants,etc. And
approximate interpolation methods are still used to get the
approximate ones$^{[16-18]}$. In order to obtain exact results,
people usually use exact interpolation methods to meliorate
intermediate expression swell problem arising from symbolic
computations $^{[19-23]}$. In fact, these are not approximate
numerical computations but big number computations, which are also
exact computations  and only improve intermediate expression swell
problem. Recently, Zhang et al. proposed an algorithm to recover the
exact rational number from its approximation$^{[24]}$, and built a
bridge by which exact results can be obtained by numerical
approximate computations. In this paper, we discuss how the errors
of support points affect that of coefficients of the interpolating
polynomial, thereby  present an algorithm to use approximate methods
to get exact interpolation multivariate  polynomial. The algorithm
can be carried out in parallel computers. Compared with exact
interpolation method, the efficiency of the our algorithm is higher
when the scale of problem is larger. This paper provide a way to
obtain exact results by numerical computations.

The remainder of the paper is organized as follows. Section 2 gives
a review of modified continued fraction method, by which an exact
rational number can be obtained from its approximation, and gives a
review of Kronecker product of two matrix. Section 3 first proposes
estimation of the error to ensure the exact interpolation polynomial
to be obtained, and then presents an algorithm to obtain an exact
polynomial from its approximation for univariate and multivariate
polynomial over rational number field,respectively. Section 4 gives
some experimental results. The final section makes conclusions.

\dse{2~~Preliminaries}

In general, obtaining exact polynomial by approximate computations
consists of two steps. First compute approximate polynomial within
an error control by approximate numerical methods, and then recover
the exact coefficients of the approximate polynomial by continued
fraction method. In this section, we review the main results to be
used in this paper.

A continued fraction representation of a real number $x$ is one of
the forms:
\begin{equation}
 a_0+\frac{1}{\displaystyle a_1+\frac{1}{\displaystyle a_2+\frac{1}{\displaystyle
 a_3+\cdots}}}
 ~,\end{equation}
where $a_0$ is an integer and $a_1,a_2,a_3,\cdots$ are positive
integers. One can abbreviate the above continued fraction as
$[a_0;a_1,a_2,\cdots]$. Truncating the above continued fraction
representation of a number $x$ early yields a rational number which
is in a certain sense the best possible rational approximation. We
call $[a_0;a_1,\cdots,a_n]$ the $n$-th {\it convergent} of
$[a_0;a_1,a_2,\cdots]$. By continued fraction representation, the
relation between a rational number and its approximation was
disclosed as follows$^{[24]}$.
\begin{theorem}\label{theo:Exact_from_Approximation}
Let $n_0/n_1$ be a reduced rational number and $r$ its
approximation. Assume that $n_0$,$n_1$ are  positive integers and
$N\ge\max\{n_1,2\}$. We have the continued fraction representation
$r=[b_0,b_1,\cdots,b_M]$ and $|r-n_0/n_1|<1/(2N^2)$. Then,
$n_0/n_1=[b_0,b_1,\cdots,b_L]$, where either $M=L$ or $M>L$. In the
case of $M>L$, for any positive integer $L<t\le M$, the denominator
of rational number $g=[b_0,b_1,\cdots,b_t]$ is greater than $N$.
\end{theorem}

Based on the theorem above, an algorithm for obtaining the exact
number was designed as follows$^{[24]}$:
\newcounter{num}
\begin{algorithm}\label{alg:Exact_from_Approximation}
Input: a nonnegative floating-point number $r$ and a positive number $N$;\\
Output: a rational number $b$.
\begin{list}{Step \arabic{num}:}{\usecounter{num}\setlength{\rightmargin}{\leftmargin}}
\item Set $i=0$,  $h_{-1}=1$, $h_{-2}=0$, $k_{-1}=0$,
$k_{-2}=1$, and convert $r$ to a rational number and assign it to
$tem$;
\item Get integral part of $tem$ and assign it to $a$, assign
its remains to $b$;
\item Compute $h_i=a*h_{i-1}+h_{i-2}$ and $k_i=a*k_{i-1}+k_{i-2}$.
If $k_i>N$, then goto Step 6; \item Set $i:=i+1$; \item If $b\ne 0$,
 set $tem=\frac{1}{b}$ and goto Step 2; \item  Compute
$h_{i-1}/k_{i-1}$ and assign it to $b$;
\item return $b$.
\end{list}
\end{algorithm}

It follows from theorem \ref{theo:Exact_from_Approximation} that
once an bound $N$ on the denominator of a positive rational number
is estimated,  we can obtain the exact rational number as follows.
Compute its approximation with the error less than $1/(2N^2)$, and
then recover the rational number from its approximation by algorithm
\ref{alg:Exact_from_Approximation}.

Now, let us give a brief review on the Kronecker product of two
matrix. Denoted by $\mathbf{M}_{m,n}(\mathbb{F})$ the set of all $m$
by $n$ matrices over field $\mathbb{F}$, and abbreviate
$\mathbf{M}_{n,n}(\mathbb{F})$ to $\mathbf{M}_{n}(\mathbb{F})$. The
Kronecker product of $\mathbf{A}=[a_{i,j}]\in
\mathbf{M}_{m,n}(\mathbb{F})$ and $\mathbf{B} \in
\mathbf{M}_{p,q}(\mathbb{F})$ is denoted by $\mathbf{A}\otimes
\mathbf{B}$ and is defined to be the block matrix
\begin{equation}
\mathbf{A}\otimes\mathbf{B}=\left(\begin{array}{ccc}a_{11}\mathbf{B}&\cdots
&a_{1n}\mathbf{B}\\ \vdots&\ddots&\vdots\\
a_{m1}\mathbf{B}&\cdots&a_{mn}\mathbf{B}
\end{array}\right)\in\mathbf{M}_{mp,nq}(\mathbb{F}).
\end{equation}

The Kronecker products has many properties$^{[25]}$. Here we mention
two properties, one of which is as follows:
\begin{theorem}[mixed-product]\label{mixed_product}
Let $\mathbf{A}\in\mathbf{M}_{m,n}(\mathbb{F})$,
$\mathbf{B}\in\mathbf{M}_{p,q}(\mathbb{F})$,
$\mathbf{C}\in\mathbf{M}_{n,k}(\mathbb{F})$, and
$\mathbf{D}\in\mathbf{M}_{q,r}(\mathbb{F})$. Then
$(\mathbf{A}\otimes\mathbf{B})(\mathbf{C}\otimes\mathbf{D})=\mathbf{AC}\otimes\mathbf{BD}$.
\end{theorem}
Another property is concerned with the eigenvalues of the Kronecker
product of two square complex matrices.
\begin{theorem}\label{eigenvalues_two_matrices}
Let $\mathbf{A}\in\mathbf{M}_n$, and $\mathbf{B}\in\mathbf{M}_m$.
Denote by $\sigma(\mathbf{A})$ the all eigenvalues of matrix
$\mathbf{A}$. If $\lambda\in\sigma(\mathbf{A})$ and
$\mathbf{x}\in\mathbb{C}^n$ is a corresponding eigenvector of
$\mathbf{A}$, and if $\mu\in\sigma(\mathbf{B})$ and
$\mathbf{y}\in\mathbb{C}^m$ is a corresponding eigenvector of
$\mathbf{B}$, then
$\lambda\mu\in\sigma(\mathbf{A}\otimes\mathbf{B})$ and
$\mathbf{x}\otimes\mathbf{y}\in\mathbb{C}^{nm}$ is a corresponding
eigenvector of $\mathbf{A}\otimes\mathbf{B}$. Every eigenvalues of
$\mathbf{A}\otimes\mathbf{B}$ arise as such a product of eigenvalues
of $\mathbf{A}$ and $\mathbf{B}$.
\end{theorem}

From theorem \ref{eigenvalues_two_matrices} and the fact that the
determinant of any matrix is equal to the product of its
eigenvalues, it follows that$^{[26]}$.
\begin{theorem}
Let $\mathbf{A}\in\mathbf{M}_n$ and $\mathbf{B}\in\mathbf{M}_m$ be
given. Then
$\det(\mathbf{A}\otimes\mathbf{B})=(\det\mathbf{A})^m(\det\mathbf{B})^n$.
Thus, $\mathbf{A}\otimes\mathbf{B}$ is nonsingular if and only if
both $\mathbf{A}$ and $\mathbf{B}$ are nonsingular.
\end{theorem}

In this paper, we need to solve an equation whose coefficient matrix
is the Kronecker product.  The following theorem shows us
 how to solve the problem$^{[25]}$.
\begin{theorem}\label{theo:kronecker_equation}
Let $\mathbb{F}$ denote a field. Matrices
$\mathbf{A}\in\mathbf{M}_{m,n}(\mathbb{F})$,
$\mathbf{B}\in\mathbf{M}_{q,p}(\mathbb{F})$, and
$\mathbf{C}\in\mathbf{M}_{m,q}(\mathbb{F})$ are given and assume
$\mathbf{X}\in\mathbf{M}_{n,p}(\mathbb{F})$ to be unknown. With
matrix $\mathbf{X}$, associate the vector $\vect{\mathbf{X}}\in
\mathbb{F}$ defined by
$$\vect{ \mathbf{X}}=[a_{11},\cdots
a_{n1},a_{12},\cdots,a_{n2},\cdots,a_{1p},\cdots,a_{np}]^T.$$ Then,
the following equation:
\begin{equation}\label{equ:kronecker_equ}
(\mathbf{B}\otimes\mathbf{A})\vect{ \mathbf{X}}=\vect{\mathbf{C}}
\end{equation}
is equivalent to  matrix equation:
\begin{equation}\label{equ:matrix_equ}
\mathbf{AXB}^T=\mathbf{C}.
\end{equation}
\end{theorem}
Obviously, equation (\ref{equ:matrix_equ}) is equivalent to the
system of equations
\begin{equation}\left\{\begin{array}{l}
\mathbf{AY}=\mathbf{C} \\
\mathbf{BX}^T=\mathbf{Y}^T.
\end{array}\right.
\end{equation}

\dse{3~~Algorithms}

    \dsee{3.1 Univariate Interpolation Polynomial}

    Let $f(x)=\sum_{i=0}^na_ix^i$ be an univariate rational polynomial
with degree $n$. Its approximate support points are $\{(x_i,f_i)\}$
for $i=0,1,\cdots,n$, which means that
$$f(x_i)=f_i,\textrm{   for } i=0,1,\cdots,n$$
where $x_i$ are called interpolation nodes and values $f_i$
interpolation datum. We can construct polynomial $f(x)$ from its
support points by interpolation method. Polynomial interpolation is
a classical numerical method. It is studied very well for univariate
polynomials. In general, interpolation problem is essentially to
solve the following equation:
\begin{equation}\label{equ:uni_interpolation}
\left(\begin{array}{ccccc}1&x_0&x_0^2&\cdots&x_0^n\\
1&x_1&x_1^2&\cdots&x_1^n\\
\cdots&\cdots&\cdots&\cdots&\cdots\\
1&x_n&x_n^2&\cdots&x_n^n
\end{array}\right)\left(\begin{array}{c}a_0\\a_1\\a_2\\ \vdots\\a_n \end{array}
\right)=\mathbf{U}_{n+1}(x_0,\cdots,x_n)\mathbf{a}=\left(\begin{array}{c}f_0\\f_1\\f_2\\
\vdots\\f_n \end{array} \right),
\end{equation}
where
\begin{equation}
\mathbf{U}_{n+1}(x_0,\cdots,x_n)=\left(\begin{array}{ccccc}1&x_0&x_0^2&\cdots&x_0^n\\
1&x_1&x_1^2&\cdots&x_1^n\\
1&x_2&x_2^2&\cdots&x_2^n\\
1&x_3&x_3^2&\cdots&x_3^n\\
\cdots&\cdots&\cdots&\cdots&\cdots\\
1&x_n&x_n^2&\cdots&x_n^n
\end{array}
\right),\ \ \mathbf{a}=\left(\begin{array}{c}a_0\\
a_1\\
a_2\\
a_3\\
\cdots\\
a_n
\end{array}
\right),
\end{equation}
Matrix $\mathbf{U}_{n+1}(x_0,\cdots,x_n)$ is called Vandermonde
matrix and its determinant
$\mathbf{V}_{n+1}(x_0,\cdots,x_n)=\det\mathbf{U}_{n+1}(x_0,\cdots,x_n)=\prod_{0\le
i<j\le n}(x_j-x_i)$ is called Vandermonde determinant. Let
$\mathbf{D}_n(i,j)$ denote the determinant of the submatrix of
matrix $\mathbf{U}_{n+1}(x_0,\cdots,x_n)$ resulting from deletion of
row $i+1$ and column $j+1$. $\mathbf{D}_n(i,j)$ is often called
generalized Vandermonde determinant. How do we compute it? The
follow theorem shows us do it$^{[27]}$.
\begin{theorem}\label{theo:generalize_vandermonde}
Let $\delta_k(x_1,x_2,\cdots,x_n)=\sum_{1\le i_1<i_2<\cdots <i_k\le
n}x_{i_1}x_{i_2}\cdots x_{i_k}$ be elementary symmetric polynomial
with degree $k$. Then $$
\mathbf{D}_n(i,j)=\delta_{n-j}(x_0,\cdots,x_{i-1},x_{i+1},\cdots,x_n)
\mathbf{V}_n(x_0,\cdots,x_{i-1},x_{i+1},\cdots,x_n).$$
\end{theorem}

From theorem \ref{theo:generalize_vandermonde}, it follows that the
solution of equation (\ref{equ:uni_interpolation})
\begin{equation}\label{equ:coeff}
a_j=\frac{\sum_{i=0}^{n}(-1)^{i+j}\mathbf{D}_n(i,j)f_i}{\mathbf{V}_{n+1}(x_0,x_1,\cdots,x_n)}.
\end{equation}

Theoretically, we can compute the interpolation polynomial after
choosing $n+1$ distinct points $x_0,x_1,\cdots,x_n$ and then
obtaining their corresponding exact interpolation values
$f_0,f_1,\cdots,f_n$. However, in practice, we often get the
approximate function values of $f(x)$, denoted by
$\tilde{f}_0,\tilde{f}_1,\cdots,\tilde{f}_n$. So an approximate
interpolation polynomial $\tilde{f}(x)=\sum_{j=0}^n\tilde{a}_jx^j$
is only produced. The following theorem gives an error estimation of
coefficients of the approximate interpolation polynomial.
\begin{theorem}\label{theo:erro_coeff}
Let $\varepsilon=\max\{|f_j-\tilde{f}_j|,j=0,\cdots,n\}$,
$\lambda_x=\min\{|x_j-x_i|,i\ne j\}$ and $M=\max\{1,
\max\{|x_j|,j=0,\cdots,n\}\}$.
\begin{equation}
\max\{|a_j-\tilde{a}_j|,j=0,\cdots,n\}<\frac{\varepsilon}{\lambda_x^n}(n+1){n\choose
\lfloor n/2\rfloor }M^{n},
\end{equation}
where $\lfloor x\rfloor$ stands for the greatest integer which is
less than or equal to $x$, and ${n\choose \lfloor n/2\rfloor }$ the
number of $\lfloor n/2\rfloor$-combinations of an $n$-element set.
\end{theorem}

Before giving the proof of theorem \ref{theo:erro_coeff}, we
introduce two lemmas:
\begin{lemma}
Let $\delta_k(x_1,x_2,\cdots,x_n)=\sum_{1\le i_1<i_2<\cdots <i_k\le
n}x_{i_1}x_{i_2}\cdots x_{i_k}$ be elementary symmetric polynomial
with degree $k$. Then
\begin{equation}\label{equ:elementary_polynomial}
\sum_{j=0}^n\delta_k(\overbrace{x_0,x_1,\cdots,x_{j-1},x_{j+1},\cdots,x_n}^{n\;
points})=(n+1-k)\delta_k(\overbrace{x_0,x_1,\cdots,x_j,\cdots,x_n}^{n+1\;
points}).
\end{equation}
\end{lemma}
{\bf Proof:}  Every term $x_{i_0}x_{i_1}\cdots x_{i_{k-1}}$ of
$\delta_k(x_0,x_1,\cdots,x_j,\cdots,x_n)$ does not appear in
$\delta_k(x_0,x_1,\cdots,x_{i-1},x_{i+1},\cdots,x_n)$ with $i=i_l$
for $ l=0,1,\cdots,k-\nobreak 1$, and appears in the other cases.
Therefore, term $x_{i_0}x_{i_1}\cdots x_{i_{k-1}}$ appears $(n+1-k)$
times on the left hand side of equation
(\ref{equ:elementary_polynomial}). So, the lemma is finished.

Another lemma is concerned with binomial coefficients as
follows$^{[28]}$:
\begin{lemma}\label{Unimodal_binomial}
For $n$ a positive integer, the largest of the binomial coefficients
$${n \choose 0},{n\choose 1},{n\choose 2},\cdots, {n\choose n}$$
is $n\choose {\lfloor n/2\rfloor}$, where $\lfloor x\rfloor$ stands
for the greatest integer which is less than or equal to $x$.
\end{lemma}
Now we turn to give the proof of theorem \ref{theo:erro_coeff}.\\
{\bf Proof:} From equation (\ref{equ:coeff}), we have
\begin{eqnarray*}
&&|a_j-\tilde{a}_j|=\left|\frac{\sum_{i=0}^{n}(-1)^{i+j}\mathbf{D}_n(i,j)(f_i-\tilde{f}_i)}{\mathbf{V}_{n+1}(x_0,x_1,\cdots,x_n)}\right|\\
&\le&\sum_{i=0}^n\frac{|\mathbf{D}_n(i,j)||(f_i-\tilde{f}_i)|}{|\mathbf{V}_{n+1}(x_0,x_1,\cdots,x_n)|}
\le
\varepsilon\sum_{i=0}^n\frac{|\mathbf{D}_n(i,j)|}{|\mathbf{V}_{n+1}(x_0,x_1,\cdots,x_n)|}\\
&\le&\varepsilon\sum_{i=0}^n\frac{|\delta_{n-j}(x_0,\cdots,x_{i-1},x_{i+1},\cdots,x_n)|
|\mathbf{V}_n(x_0,\cdots,x_{i-1},x_{i+1},\cdots,x_n)|}{|\mathbf{V}_{n+1}(x_0,x_1,\cdots,x_n)|}\\
&\le&\frac{\varepsilon}{\lambda_x^n}\sum_{i=0}^n|\delta_{n-j}(\overbrace{x_0,\cdots,x_{i-1},x_{i+1},\cdots,x_n}^{n\;
values})|,
\end{eqnarray*}
noticing equation (\ref{equ:elementary_polynomial}) yields
\begin{eqnarray*}
|a_j-\tilde{a}_j|&\le&\frac{\varepsilon}{\lambda_x^n}(n+1-n+j)\delta_{n-j}(\overbrace{|x_0|,\cdots,|x_{i-1}|,|x_i|,|x_{i+1}|,\cdots,|x_n|}^{n+1\; values})\\
&\le&\frac{\varepsilon}{\lambda_x^n}(j+1){n+1\choose n-j}M^{n-j}\\
&=&\frac{\varepsilon}{\lambda_x^n}(j+1)\frac{(n+1)!}{(n-j)!(j+1)!}M^{n-j}\\
&=&\frac{\varepsilon}{\lambda_x^n}(n+1)\frac{n!}{(n-j)!j!}M^{n-j}.
\end{eqnarray*}
From lemma \ref{Unimodal_binomial}, it follows that
\begin{eqnarray*}
|a_j-\tilde{a}_j|&\le&
\frac{\varepsilon}{\lambda_x^n}(n+1)\frac{n!}{(n-\lfloor
n/2\rfloor)!\lfloor n/2\rfloor !}M^{n-j}\\
&\le &\frac{\varepsilon}{\lambda_x^n}(n+1)\frac{n!}{(n-\lfloor
n/2\rfloor)!\lfloor n/2\rfloor !}M^{n}\\
&=&\frac{\varepsilon}{\lambda_x^n}(n+1){n\choose \lfloor n/2\rfloor
}M^{n}.
\end{eqnarray*}
The proof is completed.

And now, let us discuss how to recover the exact polynomial from its
approximate interpolation polynomial.

Let $n$ denote the degree of $f(x)$ and $N$  an upper bound of
absolute values of denominators of its coefficients. Note that
$f(x)$ is unknown, so we should estimate an upper bound $N$ and its
degree $n$ in advance. We may calculate them by the form of the
given expression of polynomial $f(x)$, and should obtain as less
bound as possible. The less bound we obtain, the less the amount of
computation is for obtaining approximate interpolation polynomial.
Once an upper bound $N$ and $n$ are gotten, we choose $n+1$
interpolate nodes $x_0,x_1,\cdots,x_n$ and calculate
\begin{equation}\label{equ:error_control}
\varepsilon=\frac{\lambda_x^n}{2(n+1){n\choose {\lfloor
n/2\rfloor}}M^nN^2}.
\end{equation}
Then, compute the values $\tilde{f}_i\approx f(x_i)$ for
$i=0,1,\cdots,n$ with an error less than $\varepsilon$. By
interpolation method, we compute the approximate interpolation
polynomial $\tilde{f}(x)$ with coefficient error less than
$1/(2N^2)$. Finally, use algorithm
\ref{alg:Exact_from_Approximation} to obtain the exact polynomial
$f(x)$ from its approximate polynomial $\tilde{f}(x)$. In summary,
the algorithm is as follows:
\begin{algorithm}
Input: an expression $f(x)$ which result is a polynomial; \\
Output: a polynomial $g(x)$ such that $g(x)=f(x)$.\\
\begin{list}{Step \arabic{num}:}{\usecounter{num}\setlength{\rightmargin}{\leftmargin}}
\item By the structure of the expression, estimate an upper bound on the degree of
$f(x)$ and an upper bound on absolute values of denominators of its
coefficients, denoted by $n$ and $N$ respectively;
\item Choose distinct points $x_0,x_1,\cdots,x_n$; Compute
$\lambda_x=\min\{|x_j-x_i|,i\ne j\}$ and
$M=\max\{1,\max\{|x_i|,i=0,\cdots,n\}\}$.
\item Compute $\varepsilon $ in formula (\ref{equ:error_control});
\item  By numerical method, approximately compute the values of $f(x)$ at the points
$x_0,x_1,\cdots,x_n$ with an error less than $\varepsilon$ and
denote the corresponding values by $\tilde{f}_i\approx f(x_i)$, for
$i=0,1,\cdots,n$ ;
\item
By interpolate method, obtain approximate interpolation polynomial
$\tilde{f}(x)$ ;
\item Call algorithm \ref{alg:Exact_from_Approximation} to recover
the exact coefficients from the coefficients of $\tilde{f}(x)$ one
by one. Denote the exact polynomial by $g(x)$;
\item return $g(x)$.
\end{list}
\end{algorithm}

The correctness of the algorithm above is shown as follows: From
step 1 to step 4, we obtain the interpolate datum $\tilde{f}_i$ with
an error less than $\varepsilon$. By theorem \ref{theo:erro_coeff},
The errors of the coefficients of the approximate polynomial are all
less than $1/(2N^2)$, which ensure the exact coefficients of the
polynomial to obtain from its approximation by algorithm
\ref{alg:Exact_from_Approximation}.

    \dsee{3.2 Multivariate Interpolation Polynomial}
    For simplicity, we first consider the bivariate interpolate problem,
and then generalize the results to the case of multivariate
interpolation. Let $f(x,y)=\sum_{i,j}a_{ij}x^iy^j$ be a polynomial
with rational coefficients, and $n$,  $m$ be the bounds on the
degree of $f(x,y)$ in $x$, $y$ respectively. We choose the
interpolation nodes $(x_i , y_j)$ ($i= 0, \cdots, n$; $j = 0, \cdots
,m$), and obtain the values of $f(x,y)$, denoted by
$f_{ij}\in\mathbb{R}$ ($i = 0, \cdots, n; j = 0,\cdots,m$). The set
of monomials is ordered as follows:
\begin{eqnarray*}
&&\{x_iy_j|i=0,\cdots,n;j=0,\cdots,m\} \\
&&=\{1, y,\cdots, y^m, x, xy,\cdots, xy^m,\cdots, x^n, x^n y,\cdots,
x^n y^m\},
\end{eqnarray*}
and the interpolation nodes in the corresponding order is as
follows:
\begin{eqnarray*}
&&\{(x_i, y_j )| i = 0, \cdots, n; j = 0,\cdots ,m\} = \{(x_0, y_0),
(x_0, y_1),\cdots,(x_0, y_m), \\
&&(x_1, y_0), (x_1, y_1),\cdots,(x_1, y_m),\cdots, (x_n, y_0),\cdots
, (x_n, y_m)\}.
\end{eqnarray*}
Note that the above ordering  is reverse to conventional
lexicographic order. Of course it is a lexicographic order under
assumption of ordering $0>1>2>\cdots $ and $y>x$. They do not
coincide with our convention, so the above ordering is called
reverse lexicographic order in this paper. Let
$$\mathbf{A}=\left( \begin{array}{ccc}
a_{00}&\cdots& a_{0m}\\
\cdots&\cdots&\cdots \\
a_{n0}&\cdots&a_{nm}
\end{array}
 \right),\quad \mathbf{F}=\left( \begin{array}{ccc}
f_{00}&\cdots& f_{0m}\\
\cdots&\cdots&\cdots \\
f_{n0}&\cdots&f_{nm}
\end{array}
 \right),$$

the bivariate interpolate problem can be expressed as to solve the
following equation:
\begin{equation}\label{equ:tem1}
\mathbf{M}\vect{\mathbf{A}^T}=\vect{\mathbf{F}^T},
\end{equation}
where the coefficients matrix
$\mathbf{M}=\mathbf{U}_x\otimes\mathbf{U}_y$. $\mathbf{U}_x$ and
$\mathbf{U}_y$ are Vandermonde matrix, i.e.,
$$\mathbf{U}_x=\left(
\begin{array}{ccccc}
1&x_0&x_0^2&\cdots&x_0^n\\
1&x_1&x_1^2&\cdots&x_1^n\\
\vdots&\vdots&\vdots&\ddots&\vdots\\
1&x_n&x_n^2&\cdots&x_n^n
\end{array}\right),\quad
\mathbf{U}_y=\left(
\begin{array}{ccccc}
1&y_0&y_0^2&\cdots&y_0^m\\
1&y_1&y_1^2&\cdots&y_1^m\\
\vdots&\vdots&\vdots&\ddots&\vdots\\
1&y_m&y_m^2&\cdots&y_m^m
\end{array}\right)
.$$

As said in section 2, in practice we often only get the approximate
interpolation values $\tilde{f}_{ij}\approx f(x_i,y_j)$. Of course,
the interpolate polynomial obtained is an approximation. The
following theorem gives an estimation error:
\begin{theorem} \label{theo:erro_coeff_multi} Set $f_{ij}=f(x_i,y_j)$ and denote by
$\tilde{f}_{ij}$ the approximation of $f_{ij}$.
$\tilde{f}(x,y)=\sum_{ij}\tilde{a}_{ij}x^iy^j$ is the interpolate
polynomial from interpolate datum $\tilde{f}_{ij}$. Let
$\varepsilon=\max\{|f_{ij}-\tilde{f}_{ij}|:i=0,\cdots,n;j=0,\cdots,m\}$,
$\lambda_x=\min\{|x_j-x_i|:i\ne j\}$,
$\lambda_y=\min\{|y_j-y_i|:i\ne j\}$ and
$M_x=\max\{1,\max\{|x_j|:j=0,\cdots,n\}\}$,
$M_y=\max\{1,\max\{|y_j|:j=0,\cdots,m\}\}$. Then
$$\max_{i,j}|a_{ij}-\tilde{a}_{ij}|\le\frac{(m+1)(n+1)M_x^nM_y^m}{\lambda_x^n\lambda_y^m}{n\choose \lfloor n/2\rfloor}{m\choose \lfloor m/2\rfloor}\varepsilon. $$
\end{theorem}
{\bf Proof}: Let $\mathbf{\tilde{A}}=(\tilde{a}_{ij})$ and
$\mathbf{\tilde{F}}=(\tilde{f}_{ij})$. From equation
(\ref{equ:tem1}), it holds that
$$\mathbf{M}\vect{(\mathbf{\tilde{A}}-\mathbf{A})^T}=\vect{(\mathbf{\tilde{F}}-\mathbf{F})^T},$$
by theorem \ref{theo:kronecker_equation}, the above equation is
equivalent to the following equation:
$$\mathbf{U}_y\mathbf{(\mathbf{\tilde{A}}-\mathbf{A})^T}\mathbf{U_x}^T=(\mathbf{F}-\mathbf{\tilde{F}})^T.$$
Thus, it is equivalent to
\begin{subequations}
\begin{eqnarray}
&&\mathbf{U}_y\mathbf{Z}=(\mathbf{F}-\mathbf{\tilde{F}})^T \label{equ:a}\\
&&\mathbf{U_x}(\mathbf{\tilde{A}}-\mathbf{A})=\mathbf{Z}^T
\label{equ:b}
\end{eqnarray}
\end{subequations}
Where $\mathbf{Z}=(z_{ij})$. Matrix equation (\ref{equ:a}) is
equivalent to
\begin{equation}
\mathbf{U}_y\mathbf{Z}_{.i}=(\mathbf{F}_{i.}-\mathbf{\tilde{F}}_{i.})^T,
\quad i=1,\cdots m+1
\end{equation}
where $\mathbf{Z}_{.i}$ stands for the $i$-th column of $\mathbf{Z}$
and $\mathbf{F}_{i.}$ the $i$-th row of matrix $\mathbf{F}$. By
theorem \ref{theo:erro_coeff}, for every $i$, it holds that
$$\max_{j=0}^m|z_{ji}|<\frac{\varepsilon}{\lambda_y^m}(m+1){m\choose
\lfloor m/2\rfloor }M_y^{m}.$$ Hence, we conclude that
$$\max_{i,j}|z_{ji}|<\frac{\varepsilon}{\lambda_y^m}(m+1){m\choose
\lfloor m/2\rfloor }M_y^{m}.$$ Let
$$\delta=\frac{\varepsilon}{\lambda_y^m}(m+1){m\choose \lfloor
m/2\rfloor }M_y^{m},$$ argue equation (\ref{equ:b}) in the same way
as do above, we deduce that
$$\max_{i,j}|a_{ij}-\tilde{a}_{ij}|\le\frac{(m+1)(n+1)M_x^nM_y^m}
{\lambda_x^n\lambda_y^m}{n\choose \lfloor n/2\rfloor}{m\choose
\lfloor m/2\rfloor}\varepsilon. $$ The proof is finished.

Now, After having a relation between error of approximate
coefficients and that of interpolate datum, we establish an
algorithm to obtain an exact polynomial from its approximation as
follows:
\begin{algorithm} \label{alg:Exactbivariate}
Input: an expression $f(x,y)$ which result is a polynomial; \\
Output: a polynomial $g(x,y)$ such that $g(x,y)=f(x,y)$.
\begin{list}{Step \arabic{num}:}{\usecounter{num}\setlength{\rightmargin}{\leftmargin}}
\item By the structure of the expression, estimate an upper bounds on the degrees of
$f(x,y)$ in $x$ and $y$, denoted by $n$ and $m$ respectively.
Estimate an upper bound on absolute values of denominators of its
coefficients, denoted by $N$;
\item Choose interpolate nodes $(x_i,y_j)$,($i=0,1,\cdots,n;j=0,1,\cdots,m$). Compute
$\lambda_x=\min_{j\ne i}\{|x_j-x_i|\}$, $\lambda_y=\min_{j\ne
i}\{|y_j-y_i|\}$ and $M_x=\max\{1,\max_{i=0}^n|x_i|\}$,
$M_y=\max\{1,\max_{i=0}^m|y_i|\}$.
\item Compute $\varepsilon $:
$$\varepsilon=\frac{\lambda_x^n\lambda_y^m}{2(n+1)(m+1){n\choose {\lfloor
n/2\rfloor}}{m\choose {\lfloor m/2\rfloor}}M_x^nM_y^mN^2}; $$
\item  By numerical method, compute the approximate values of $f(x,y)$ at the points
$(x_i,y_j)$($i=0,1,\cdots,n;j=0,1,\cdots,m$) with an error less than
$\varepsilon$, and denote the corresponding values by
$\tilde{f}_{ij}\approx f(x_i,y_j)$,
($i=0,1,\cdots,n;j=0,1,\cdots,m$) ;
\item
By interpolate method, obtain approximate interpolation polynomial
$\tilde{f}(x,y)$;
\item Call algorithm \ref{alg:Exact_from_Approximation} to recover
the exact coefficients from the coefficients of $\tilde{f}(x,y)$ one
by one. Denote the exact polynomial by $g(x,y)$;
\item return $g(x,y)$.
\end{list}
\end{algorithm}

As for generalization of the above results to the case of
multivariate polynomials $r>2$, we assert that the situation is
completely analogous to the bivariate case.

Let $f(X_1,\cdots,X_r)=\sum a_{i_1\cdots i_r}X_1^{i_1}\cdots
X_r^{i_r}$ be a polynomial in $X_1,\cdots,X_r$, and let $n_i$ be a
bound on the degree of $f(X_1,\cdots, X_r )$ in $X_i$ ($i =
1,\cdots, r$ ). Denotes by $N$ an upper bound on absolute values of
denominators of coefficients of polynomial $f(X_1,\cdots,X_r)$.
Choose interpolation nodes $(x_{1i_1},x_{2i_2}\cdots,x_{ri_r})$
($i_k=0,1,\cdots,n_k$), where $k=1,\cdots,r$. Set $f_{i_1i_2\cdots
i_r}=f(x_{1i_1},x_{2i_2},\cdots,x_{ri_r})$ and denote by
$\tilde{f}_{i_1i_2\cdots i_r}$ the approximation of $f_{i_1i_2\cdots
i_r}$, by which we compute an approximate interpolation polynomial
$\tilde{f}(X_1,X_2,\cdots,X_r)=\sum\tilde{a}_{i_1i_2\cdots
i_r}X_1^{i_1}X_2^{i_2}\cdots X_r^{i_r} $. Thus we have the following
theorem:
\begin{theorem}  \label{theo:erro_coeff_multi2}
Let $\varepsilon=\max_{i_1,i_2,\cdots i_r}|f_{i_1i_2\cdots
i_r}-\tilde{f}_{i_1i_2\cdots i_r}|$, $\lambda_k=\min_{i\ne
j}|x_{kj}-x_{ki}|$, and $M_k=\max\{1,\max_{j=0}^{n_k}|x_{kj}|\}$.
Then
\begin{equation}\label{equ:conclusion_theorem_multi}
\max_{i_1,i_2,\cdots,i_r}|a_{1i_1,2i_2,\cdots,ri_r}-\tilde{a}_{1i_1,2i_2,\cdots,ri_r}|\le
\varepsilon\prod_{k=1}^r\frac{(n_k+1)M_k^{n_k}{n_k\choose \lfloor
n_k/2\rfloor}} {(\lambda_k^{n_k})}.
\end{equation}
\end{theorem}
{\bf Proof}: By the Reverse Lexicographic order, the interpolation
problem comes down to solving the following equation:
$$(\mathbf{U}_{X_1}\otimes\mathbf{U}_{X_2}\cdots\otimes\mathbf{U}_{X_r})\mathbf{a}=\mathbf{F},$$
where $\mathbf{a}$ is a column vector consisting of elements
$a_{i_1i_2\cdots i_r}$ in the reverse lexicographic order and
$\mathbf{F}$ is a column vector consisting of elements
$f_{i_1i_2\cdots i_r}$ in the reverse lexicographic order. Thus we
have the following equation:
\begin{equation}\label{equ:multi_interpolation}
(\mathbf{U}_{X_1}\otimes\mathbf{U}_{X_2}\cdots\otimes\mathbf{U}_{X_r})(\mathbf{a}-\mathbf{\tilde{a}})=(\mathbf{F}-\mathbf{\tilde{F}}),
\end{equation}
where $\mathbf{\tilde{a}}$ and $\mathbf{\tilde{F}}$ are column
vectors respectively consisting of elements $\tilde{a}_{i_1i_2\cdots
i_r}$ and $\tilde{f}_{i_1i_2\cdots i_r}$ in the reverse
lexicographic order. Equation (\ref{equ:multi_interpolation}) is
equivalent to  the following equation:
$$((\mathbf{U}_{X_1}\otimes\mathbf{U}_{X_2}\cdots \otimes\mathbf{U}_{X_{r-1}})\otimes\mathbf{U}_{X_r})(\mathbf{a}-\mathbf{\tilde{a}})=
(\mathbf{F}-\mathbf{\tilde{F}}).$$ Hence,  by recursion and as did
in the proof of theorem \ref{theo:erro_coeff_multi}, we can deduce
that inequality (\ref{equ:conclusion_theorem_multi}) holds. The
proof is finished.

Based on the above theorem, algorithm \ref{alg:Exactbivariate} is
generalized to the case of multivariate interpolation as follows:
\begin{algorithm}\label{alg:Exactmultivariate}
Input: an expression $f(X_1,\cdots,X_r)$ which result is a polynomial; \\
Output: a polynomial $g(X_1,\cdots,X_r)$ such that
$g(X_1,\cdots,X_r)=f(X_1,\cdots,X_r)$.
\begin{list}{Step \arabic{num}:}{\usecounter{num}\setlength{\rightmargin}{\leftmargin}}
\item By the structure of the expression, estimate bounds on the degrees of
$f(X_1,\cdots,X_r)$ in $X_i$ , denoted by $n_i$($i=1,\cdots,r$) .
Estimate an upper bound on absolute values of denominators of its
coefficients, denoted by $N$;
\item Choose interpolation nodes $(x_{1i_1},\cdots,x_{ri_r})$,($i_k=0,1,\cdots,n_k;k=0,1,\cdots,r$). Compute
$\lambda_k=\min_{j\ne i}\{|x_{kj}-x_{ki}|\}$ and
$M_k=\max\{1,\max_{i=0}^n|x_{ki}|\}$ for $k=1,\cdots,r$.
\item Compute $\varepsilon $:
\begin{equation}\label{equ:multi_error_control}
\varepsilon=\frac{\prod_{k=1}^r(\lambda_k^{n_k})}{2N^2\prod_{k=1}^r(n_k+1)M_k^{n_k}{n_k\choose
\lfloor n_k/2\rfloor}};
\end{equation}
\item  By numerical method, compute the approximate values of $f(X_1,\cdots,X_r)$ at the points
$(x_{1i_1},\cdots,x_{ri_r})$($i_k=0,1,\cdots,n_k;k=0,1,\cdots,r$)
with an error less than $\varepsilon$ and denote the corresponding
values by $\tilde{f}_{i_1\cdots i_r}\approx
f(x_{1i_1},\cdots,x_{ri_r})$, ($i_k=0,1,\cdots,n_k;k=0,1,\cdots,r$)
;
\item
By interpolate method, obtain approximate interpolation polynomial
$\tilde{f}(X_1,\cdots,X_r)$;
\item Call algorithm \ref{alg:Exact_from_Approximation} to recover
the exact coefficients from the coefficients of
$\tilde{f}(X_1,\cdots,X_r)$ one by one. Denote the exact polynomial
by $g(X_1,\cdots,X_r)$;
\item return $g(X_1,\cdots,X_r)$.
\end{list}
\end{algorithm}

The correctness of the above algorithm  is ensured by theorem
\ref{theo:erro_coeff_multi} and theorem
\ref{theo:Exact_from_Approximation}. The proof can be given
similarly to that in algorithm \ref{alg:Exactbivariate}.

\dse{4~~Experimental Results} Our algorithms are implemented in
\emph{Maple}. The following examples run in the platform of Maple11
and PIV3.0G,512M RAW. The first three examples illuminate how to
obtain exact interpolation polynomials. Example 4 tests our
algorithm  for more variables and higher degree interpolation
multivariate polynomials.

\textbf{Example 1} Let $f(x)$ be an unknown univariate rational
polynomial. Assume that we know in advance the degree of $f(x)$:
$n=8$ and an upper bound of absolute values of denominators of its
coefficients $N=181$. According to theorem\ref{theo:erro_coeff},
computing exact interpolation polynomial $f(x)$ as follows: Choose
floating-point interpolation
nodes$[0.5001,1.5003,3.1201,2.314,4.02,5.23,6,6.8,\\7.2001]$;
Calculate $\lambda_{x}=0.4001$, M=7.2001, Compute
$\varepsilon=0.2202\times10^{-17}$. Compute the approximate
interpolate datum $\tilde{f}_i\approx f(x_i)$ such that
$|f(x_{i})-\tilde{f_{i}}|<\varepsilon$, for $ i=0,\cdots,8$, and use
interpolation method to obtain approximate interpolation polynomial
$\tilde{f}(x)$:
\begin{eqnarray*}
&&.33333333333333327598x^{8}-.027777777777775996307x^{7}
+.66666666666664376920x^{6}\\&&
-.072222222222064098893x^{5}+1.9583333333326967612x^{4}+1.7500000000015123646x^{3}\\&&
-.24166666666870050587x^{2}+.17500000000137107868x
+1.4999999999996690437.
\end{eqnarray*}
Then we recover the exact coefficients of interpolation polynomial
by algorithm \ref{alg:Exact_from_Approximation}:
$$
\emph{f(x)}=\frac{1}{3}x^{8}-\frac{1}{36}x^{7}+\frac{2}{3}x^{6}-\frac{13}{180}x^{5}+\frac{47}{24}x^{4}+\frac{7}{4}x^{3}-\frac{29}{120}x^{2}+\frac{7}{40}x+\frac{3}{2}.
$$

\textbf{Example 2} Let $f(x,y)$ be an unknown bivariate rational
polynomial. Assume that we  know in advance the degrees of $f(x,y)$
in $x$ and $y$: $n=3,m=3$ , and an upper bound of absolute values of
denominators of its coefficients $N=13$. According to theorem
\ref{theo:erro_coeff_multi}, computing exact interpolation
polynomial $f(x,y)$ as follows: Choose floating-point interpolation
nodes $X=[0.1,0.5,1.2,1.3]$, $Y=[0.2,0.8,2.1,2.6]$, and calculate
$\lambda_{x}=0.1,\lambda_{y}=0.5, M_{x}=1.3,M_{y}=2.6$; Compute
$\varepsilon=0.6659\times10^{-10}$. Compute the approximate
interpolate datum $\tilde{f}_{ij}$ such that
$|f(x_{i},y_{j})-\tilde{f}_{ij}|<\varepsilon$, for
$i=0,\cdots,3,j=0,\cdots,3$, and use interpolation method to obtain
approximate interpolation polynomial $\tilde{f}(x,y)$ as follows:
\begin{eqnarray*}
&&-.2500000301x^{3}y^{2}-.2500000111x^{2}y^{3}+.08333326696x^{2}y-.08333336318xy^{2}\\&&
-.1666666674y^{3}-.2499999755x^{2}+.2500000031y^{2}-.5000000114x-.5000000029y.
\end{eqnarray*}
Then we recover the exact coefficients of interpolation polynomial
by algorithm \ref{alg:Exact_from_Approximation}:
$$
f(x,y)=-\frac{1}{4}x^{3}y^{2}-\frac{1}{4}x^{2}y^{3}+\frac{1}{12}x^{2}y-\frac{1}{12}xy^{2}
-\frac{1}{6}y^{3}-\frac{1}{4}x^{2}+\frac{1}{4}y^{2}-\frac{1}{2}x-\frac{1}{2}y.
$$

\textbf{Example 3} Let $f(x,y,z)$ be an unknown tri-variate rational
polynomial. Assume that we know in advance the degrees of $f(x,y,z)$
in $x,y,z$: $n=3,m=2,l=3$, and an upper bound of absolute values of
denominators of its coefficients $N=231$. According to theorem
\ref{theo:erro_coeff_multi2}, computing exact interpolation
polynomial $f(x,y,z)$ as follows: Choose floating-point
interpolation nodes $X=[0.5, 1.2, 1.8, 2.7],Y=[0.3, 1.5,
2.2],Z=[0.6, 1.8, 2, 2.8]$, and calculate
$\lambda_{x}=0.6,\lambda_{y}=0.7,\lambda_{z}=0.4,
M_{x}=2.7,M_{y}=2.2,M_{z}=2.8$; Compute
$\varepsilon=0.735\times10^{-10}$. Compute the approximate
interpolate datum $\tilde{f}_{ijk}$ such that
$|f(x_{i},y_{j},z_{k})-\tilde{f}_{ijk}|<\varepsilon$, for
$i=0,\cdots,3,j=0,\cdots,2,k=0,\cdots,3$, and use interpolation
method to obtain approximate interpolation polynomial
$\tilde{f}(x,y,z)$ as follows:
\begin{eqnarray*}
&&-.08333331999x^{3}y^{2}+.1666668454x^{2}y^{2}z-.004629485562xy^{2}z^2{}-.02777770742xyz^{3}\\
&&-.05555579922xy^{2}z-.3333337487xyz^{2}+.08333353340x^{2}y-.1666659870xyz\\
&& +.01388866836yz+.08333327541z^{2}+.5000001136x+.1666667565y.
\end{eqnarray*}
Then we recover the exact coefficients of interpolation polynomial
by algorithm \ref{alg:Exact_from_Approximation}:
\begin{eqnarray*}
f(x,y,z)=-\frac{1}{12}{x}^{3}{y}^{2}+\frac{1}{6}{x}^{2}{y}^{2}z-\frac{1}{216}x{y}^{2}{z}^{2}-\frac{1}{36}{x}{y}{z}^{3}-\frac{1}{18}x{y}^{2}z && \nonumber\\
-\frac{1}{3}{x}y{z}^{2}+\frac{1}{12}{x}^{2}{y}-\frac{1}{6}{x}{y}{z}+\frac{1}{72}{y}z+\frac{1}{12}{z}^{2}+\frac{1}{2}x+\frac{1}{6}y.
\end{eqnarray*}

\textbf{Example 4} This example tests algorithm
\ref{alg:Exactmultivariate} by computing determinants, and the
results are shown in table 1, where '?' represents that the
computation is not finished.  Table 1 shows that our algorithm is
more efficient than exact interpolation algorithm when scale of
problem is larger.
\begin{table}[h]
\begin{center}
\caption{\label{ta1}Comparison between our algorithm and exact
interpolation }\vskip3mm

\begin{tabular}
{|p{1.4cm}| p{1.2cm}|p{1cm}|p{2.5cm}|p{1.75cm}|p{1.75cm}|}
\hline \textbf{matrix sizes}
 &\textbf{variates}&
\textbf{terms} &\textbf{error control}
&\textbf{time(sec)}&\textbf{exact interp(sec) }\\[0.3ex]
\hline \textbf{$3\times3$} & $ 3$&$210 $&$0.1679e-26$&$1.766$&$1.531$\\
\hline \textbf{$3\times3$}&$4$ &$343$   &$0.1829e-30 $   &$5.578 $ &$3.837$ \\
\hline \textbf{$4\times4$}&$4$ &$256$   &$0.5441e-19$    &$2.078$ &$1.734$ \\
\hline \textbf{$5\times5$}&$3$ & 816  &$0.4471e-37 $   &$ $5.837 &$4.241$\\
\hline \textbf{$6\times6$}&$3$ &2774 &$0.3781e-45$    &$147.452$  &$67.731 $\\
\hline \textbf{$7\times7$}&$4$&$5657$  &$0.4731e-50$   &$ 258.129$&$347.372$ \\
\hline \textbf{$8\times8$}&$4$ &$8847$   &$0.4251e-55$    &$657.221 $&$899.372$ \\
\hline \textbf{$9\times9$}&$3 $&$ 15367$ &$0.5882e-63 $  &$2074.771 $ &$2744.305$\\
\hline \textbf{$10\times10$}&4&$23371 $ &$0.7321e-61 $    &$5721.491$  &$8737.707$\\
\hline \textbf{$11\times11$}&$4$ & $ 40721$&$0.6781e-73$   &$ 9287.527$&$?$ \\
\hline
\end{tabular}
\end{center}
\vspace{-0.15in}
\end{table}

\dse{5~~Conclusions}
 The paper presents a new method for obtaining exact interpolation multivariate polynomial
with rational coefficients from its approximate interpolation. The
exact results can be obtained by our algorithm as long as we
estimate the upper bounds on the degrees in every variable  and
upper bound $N$ on absolute values of denominators of its
coefficients. As we know, many complicate polynomials computation
can be completed by exact interpolation method, so they can also be
completed by our algorithm, i.e., by  approximately numerical
computation.

The paper proposes a way to get an exact polynomial by approximate
numerical computation. Like exact interpolation methods, this
algorithm is a parallel method. The accuracy control $\varepsilon $
in formula (\ref{equ:multi_error_control}) is an exponential
function in degrees of polynomial, and is a polynomial function in
the upper bound $1/N$. Experimental results show that our method is
more efficient than the exact interpolation methods when scale of a
problem gets larger.

%
%
%

\rfne

\rf{1} {Robert M. Corless, Mark W. Giesbrecht, et al, Towards
factoring bivariate approximate polynomials, In Proc. ISSAC 2001,
ACM press, pp.85-92.}
\rf{2} {Huang Y., Wu W.,Stetter H., and Zhi L.
Pseudofactors of multivariate polynomials. In Proc. ISSAC¡¯00(2000),
ACM Press, pp.161-168.}
 \rf{3} {Mou-Yan, Z., and Unbehausen, R.
Approximate factorization of multivariable polynomials. Signal
Proces. {\bf 14}(1988), pp.141-152.}
\rf{4} { Sasaki, T., Suzuki,
M., et al., Approximate factorization of multivariate polynomials
and absolute irreducibility testing. Japan J. Indust. Appl. Math. 8
(1991),pp.357-375.}
\rf{5} {Tateaki Sasaki, Approximate multivariate
polynomial factorization based on zero-sum relations. In Proc.
ISSAC¡¯2001, ACM Press,pp.284- 291.}
\rf{6} {Sasaki, T., Saito T.,
and Hilano, T., Analysis of approximate factorization algorithm.
Japan J. Indust. Appl. Math.9 (1992),pp.351-368.}
\rf{7} { Corless
R. M., Gianni P.M., Trager B.M. and Watt, S.M. The singular value
decomposition for polynomial systems. In International Symposium on
Symbolic and Algebraic Computation (Montreal, Canada, 1995), A
Levelt, Ed., ACM pp.195-207.}
\rf{8} {Beckermann, B., and Labahn,
G., When are two polynomials relatively prime? Journal of Symbolic
Computation 26 (1998), pp.677-689.}
 \rf{9} {Karmarka N., and
Lakshman Y. N., Approximate polynomial greatest common divisors and
nearest singular polynomials. In Proceeding of ISSAC 1996, ACM,
pp.35 -42.} \rf{10} {Robert M. Corless , Stephen M. Watt, and Lihong
Zhi, QR Factoring to Compute the GCD of Univariate Approximate
Polynomials IEEE Transactions on Signal Processing, {\bf
52}(12),(2004), pp.3394-3402.}
\rf{11} {Robert M. Corless, MarkW.
Giesbrecht, et al, Approximate polynomial decomposition. In
proceeding of ISSAC 1999, S.S. Dooley, Ed., ACM pp.213-220.}
\rf{12}
{Galligo A., and Watt S. M. A numerical absolute primality test for
bivariate polynomials In proceeding of ISSAC 1997.
W.K$\ddot{u}$chlin, Ed. ACM, pp.217-224.}
\rf{13} {Greg Reid , and
Lihong Zhi, Solving Nonlinear Polynomial System via Symbolic-Numeric
Elimination Method, In Proceedings of international conference on
polynomial system solving,(2004), pp.50-53.}
 \rf{14} {Corless, R.M.,
Giesbrecht,M.W., et al, Numerical implicitization of parametric
hypersurfaces with linear algebra. In proceeding of AISC2000, LNAI
1930, pp.174-183.} \rf{15}{ G.Ch$\grave{e}$ze,A. Galligo. From an
approximate to an exact absolute polynomial factorization. J.
Symbolic Comput., {\bf 41}(6),(2006),pp.682-696.}
\rf{16} {Erich
Kaltofen,Lakshman Yagati, Improved Sparse Mutivariate Polynomial
Interpolation Algorithms. In Symbolic Algebraic Comput. Internat.
Symp. ISSAC '88 Proc. [-17], pages 467-474, 1988.}
\rf{17} {Manocha
D,Efficient Algorithms for multipolynomial Resultant.The Computer
Journal,1993,36(5)£¬485-496.}
\rf{18} {M.Giesbrecht, G. Labahn and
W-s Lee, Symbolic-numeric Sparse Interpolation of Multivariate
Polynomials, Proceedings of ISSAC'06, Genoa, Italy, ACM Press,
(2006) 116-123.}
\rf{19} {A Marco, JJ Martinez, Parallel computation
of determinants of matrices with polynomial entries. J. Symbolic
Comput. 37(2004), No.6,749-760.}
\rf{20} {Tushar Saxena,Efficient
Variable Elimination using Resultants, Doctoral Thesis, Computer
Science Division, State University of New York at Albany. 1997.}
\rf{21}{ Yong Feng,Yaohui Li,Checking RSC Criteria for Extended
Dixon Resultant by Interpolation Method,Proceedings of 7-th
International Symposium on Symbolic and Numeric Algorithms for
Scientific Computing,Timisoara,Romania,September 25-29,2005,IEEE
Computer Society press pages 48-51.}
\rf{22} {Yong Feng,Yaohui
Li,Computing extended Dixon resultant by interpolation
method.Proceedings of international workshop on Symbolic-Numeric
Computation.Xi'an China,July 19-21,2005,pp.177-187.}
\rf{23} {Carl
de Boor, Polynomial Interpolation in Several Variables, in Studies
in Computer Science (in Honor of Samuel D. Conte), R. DeMillo and J.
R. Rice (eds.), P87¨C119, Plenum Press New York. 1994.}
\rf{24}{
Jingzhong Zhang, YongFeng, Obtaining Exact Value by Approximate
Computations. Science in China Series A: Mathematics Vol. 50, No.9,
1361-1368, Sep. 2007.}
\rf{25} {R.A., Johnson, C. R.,Topics in
Matrix Analysis, Cambridge University Press, Cambridge, 1991.}
\rf{26} {John W. BREWER, Kronecker products and matrix calculus in
system theory, IEEE Transactions on Circuits and Systems, 1978.}
\rf{27} {Xianke Zhang, Puhua Xu, Higher Algebra(second edition) (in
Chinese), Beijing: Tsinghua University Press, 2004.}
 \rf{28}{
Richard A. Brualdi, Introductory Combinatorics. North-Holland
Publishing Company, NewYork Oxford Amsterdam, 1977.}

\end{document}